%% file: main.tex
\documentclass[11pt,a4paper]{article}
\usepackage[T1]{fontenc}
\usepackage[utf8]{inputenc}
\usepackage{lmodern}
\usepackage[margin=25mm]{geometry}
\usepackage{amsmath,amssymb,amsthm,mathtools}
\usepackage{enumitem,booktabs,array}
\usepackage{algorithm}
\usepackage{microtype}
\usepackage[authoryear,round]{natbib}
\usepackage[hidelinks]{hyperref}
\hypersetup{pdftitle={Improved Bounds for Online Makespan Scheduling on Related Machines},pdfauthor={Tianqi Chen and Junwen Xiao},pdfkeywords={Scheduling; Online algorithms; Related machines; Makespan minimization; Competitive analysis}}
\newtheorem{theorem}{Theorem}[section]
\newtheorem{lemma}[theorem]{Lemma}

\theoremstyle{remark}

\DeclareMathOperator{\OPT}{OPT}

\setlist{itemsep=3pt,topsep=5pt}
\allowdisplaybreaks[1]
\newcommand{\keywords}[1]{\par\smallskip\begingroup\footnotesize\noindent\textbf{Keywords.} #1\par\endgroup}
\title{Improved Bounds for Online Makespan Scheduling\\on Related Machines}
\author{Tianqi Chen\thanks{School of Mathematical Science, Zhejiang University, Hangzhou 310027, P. R. China.\newline Email: \href{mailto:ctq@zju.edu.cn}{\nolinkurl{ctq@zju.edu.cn}}.}
  \and Junwen Xiao\thanks{School of Mathematical Science, Zhejiang University, Hangzhou 310027, P. R. China.\newline Email: \href{mailto:12435011@zju.edu.cn}{\nolinkurl{12435011@zju.edu.cn}}.}}
\date{}
\begin{document}
\maketitle
\vspace{-1.3em}
\begin{abstract}
We study online makespan minimization on related machines with known
positive speeds. Jobs arrive one at a time and must be assigned
immediately and irrevocably, without knowledge of the sizes or number of
future jobs. We give a deterministic algorithm with competitive ratio
$\frac{3+\sqrt5}{2}\exp((\sqrt5-1)/2)\approx4.857178$ for an arbitrary
number of machines, improving the upper bound of
$3+\sqrt8\approx5.8284$. We also raise the randomized lower bound from
$2$ to $e$ as the number of machines tends to infinity. Both results
improve bounds established more than twenty-five years ago.
\end{abstract}
\keywords{Scheduling; Online algorithms; Related machines; Makespan minimization; Competitive analysis}
\input{sections/01_introduction}
\input{sections/02_deterministic_algorithm}
\input{sections/03_upper_bound}
\input{sections/04_lower_bound}
\input{sections/05_conclusion}
\bibliographystyle{apalike}
\bibliography{references}
\end{document}

%% file: sections/01_introduction.tex
\section{Introduction}\label{sec:introduction}

Online makespan minimization on related machines asks how arriving jobs
should be distributed among machines of different speeds so as to keep
the maximum load small. Unequal speeds create a tension between keeping
the current makespan small and reserving capacity for future jobs.
Assigning early small jobs to fast machines can keep the current makespan
low, but consumes capacity that may be needed by later large jobs.
Assigning those small jobs to slower machines preserves fast-machine
capacity, but may already incur a large competitive ratio if the input
ends immediately. For any target makespan, a job can use only machines
fast enough to process it within that target. We therefore need to compare
the total size of large jobs with the capacity of the machines that can
process them within the target makespan.

Classical work established constant competitive ratios for related
machines, with guarantees independent of the number of machines and their
speed ratios~\citep{AspnesEtAl1997,BermanCharikarKarpinski2000}. Substantial gaps
remain between the known upper and lower bounds. We address both
directions by improving the deterministic upper bound and strengthening
the randomized lower bound.

\subsection{Model and notation}
\label{sec:model}
\enlargethispage{\baselineskip}

There are $m\ge1$ machines with known speeds
$\boldsymbol{s}=(s_1,\ldots,s_m)$, where $0<s_1\le\cdots\le s_m$.
Jobs arrive one at a time. Job $j$ has size $p_j>0$, and its
processing time on machine $i$ is $p_j/s_i$.
Every job may be assigned to any machine. Upon arrival, the entire
job must be assigned to one machine before the next job is revealed.
The assignment is irrevocable, and jobs cannot be preempted or migrated.
Zero-size jobs can be assigned arbitrarily on arrival and otherwise
ignored without affecting the competitive guarantees.

Let $I=(p_1,\ldots,p_n)$ be a finite input, and let
$I_t=(p_1,\ldots,p_t)$ denote its first $t$ jobs for $1\le t\le n$,
with $I_n=I$.
An online algorithm $\mathcal A$ knows the machine speeds before
the first arrival. Its decision for job $t$ may depend only on
$I_t$, its previous assignments, and its random choices.
The sizes of future jobs and the number of future jobs are unknown.
After $\mathcal A$ has assigned the jobs in $I_t$, let $W_i(t)$
denote the total work of machine $i$, that is, the sum of the sizes
of the jobs assigned to it.
The load of machine $i$ is $W_i(t)/s_i$, and the makespan is
$C_{\max}^{\mathcal A}(I_t)=\max_{1\le i\le m} W_i(t)/s_i$.
Let $\OPT(I_t)$ denote the offline optimum: the minimum
makespan over all assignments of the jobs in $I_t$
to the machines. The minimum is attained because there are
finitely many such assignments.
When the input, speed vector, and algorithm $\mathcal A$ are clear,
we write $C_{\max}(t)=C_{\max}^{\mathcal A}(I_t)$ and
$\OPT_t=\OPT(I_t)$.

A deterministic online algorithm $\mathcal A$ is
$\alpha$-competitive, for $\alpha\ge1$, if for every positive
speed vector $\boldsymbol{s}$ of arbitrary length and every
finite input $I$,
\[
C_{\max}^{\mathcal A}(I)
\le
\alpha\,\OPT(I).
\]

For a randomized algorithm, the input may depend on the algorithm
and its probability law but is fixed independently of its realized
random choices. All expectations are taken over the algorithm's
random choices.
A randomized online algorithm $\mathcal A$ is
$\alpha$-competitive, for $\alpha\ge1$, if for every positive
speed vector $\boldsymbol{s}$ of arbitrary length and every
fixed finite input $I$,
\[
\mathbb E[C_{\max}^{\mathcal A}(I)]
\le
\alpha\,\OPT(I).
\]

In either case, the competitive ratio is the infimum of the values
$\alpha\ge1$ for which the corresponding guarantee holds.
Since the algorithm uses only the jobs revealed so far and does
not know the number of future jobs, the guarantee also applies
to every fixed $I_t$:
$C_{\max}(t)\le\alpha\,\OPT_t$ in the deterministic case
and $\mathbb E[C_{\max}(t)]\le\alpha\,\OPT_t$
in the randomized case.

\subsection{Our results}
\label{subsec:our-results}

For an arbitrary number of related machines, Berman, Charikar,
and Karpinski~\citeyearpar{BermanCharikarKarpinski2000} obtained a deterministic competitive ratio
of $3+\sqrt8\approx5.8284$ and a randomized competitive ratio
of approximately $4.3111$.
On the lower-bound side, \citet{EpsteinSgall2000} proved a bound
of $2$ for randomized algorithms, while \citet{EbenlendrSgall2015}
established a lower bound of approximately $2.5650$ for deterministic algorithms.
These results leave substantial gaps between the upper and lower
bounds for both classes of algorithms.

Our first result improves the deterministic upper bound to
$\alpha_0$ for an arbitrary number of related machines
(Theorem~\ref{thm:analytic}), where
\[
\alpha_0
=
\frac{3+\sqrt5}{2}
\exp\!\left(\frac{\sqrt5-1}{2}\right)
\approx4.857178.
\]
Our second result establishes a randomized lower bound of $e$
as $m\to\infty$ (Theorem~\ref{thm:lb-main}).
This also gives an asymptotic deterministic lower bound of $e$.
Together, these results narrow the gap between the known upper and lower
bounds by approximately $34\%$ for deterministic algorithms and $31\%$
for randomized algorithms.
Table~\ref{tab:prior-bounds} summarizes the bounds.

\begin{table}[!htbp]
\centering
\small
\caption{Competitive ratios on related machines.
The two lower bounds of $e$ are asymptotic in the number of machines.}
\label{tab:prior-bounds}
\renewcommand{\arraystretch}{1.13}
\begin{tabular}{@{}lcccc@{}}
\toprule
 & \multicolumn{2}{c}{Deterministic} & \multicolumn{2}{c}{Randomized}\\
\cmidrule(lr){2-3}\cmidrule(lr){4-5}
 & Lower bound & Upper bound & Lower bound & Upper bound\\
\midrule
Previous results & $2.5650$ & $5.8284$ & $2$ & $4.3111$\\
This paper & $e$ & $4.857178$ & $e$ & ---\\
\bottomrule
\end{tabular}
\par\smallskip
\begin{minipage}{\linewidth}\footnotesize
\citet{EbenlendrSgall2015} and \citet{EpsteinSgall2000} give the previous deterministic and randomized lower bounds, respectively. \citet{BermanCharikarKarpinski2000} give both previous upper bounds.
\end{minipage}
\end{table}

\subsection{Related work}\label{sec:related-work}

\paragraph{Identical machines.}
Online makespan minimization has long been studied on identical parallel
machines. \citet{Graham1966} showed that list scheduling, which assigns each
arriving job to a least-loaded machine, is $(2-1/m)$-competitive.
This ratio is optimal among deterministic online algorithms for
$m\le3$~\citep{FaigleKernTuran1989}. For a general number of machines,
subsequent algorithms improved the asymptotic deterministic guarantee
below $2$. \citet{BartalEtAl1995} obtained a bound of $2-1/70\approx1.9857$, followed by
$1.945$ due to \citet{KargerPhillipsTorng1996} and $1.923$ due to \citet{Albers1999}.
\citet{FleischerWahl2000} further improved the asymptotic guarantee to
approximately $1.9201$.

On the lower-bound side, \citet{Rudin2001} established a bound of $1.88$
for deterministic algorithms, which \citet{XiaoTan2026} improved to
$1.8871$. For four machines, \citet{Janke2026} raised the lower bound
from $\sqrt3$ to $14350/8283\approx1.7325$.

The effect of randomization has also received considerable attention.
\citet{ChenVanVlietWoeginger1994} and \citet{Sgall1997} developed lower bounds for randomized
online algorithms, while \citet{Albers2002} gave a $1.916$-competitive
randomized algorithm. In these results, the arrival sequence is fixed
and the expectation is taken over the algorithm's random choices.
A different source of randomness is studied by \citet{AlbersJanke2021},
who consider a uniformly random arrival order and obtain an asymptotic
competitive ratio of $1.8478$ for a deterministic algorithm as the
number of machines grows.

\paragraph{Related machines.}
\citet{ChoSahni1980} showed that list scheduling has competitive ratio
at most $1+\sqrt{(m-1)/2}$ for $m\ge3$. For two machines, its exact
competitive ratio is $(1+\sqrt5)/2$. \citet{AspnesEtAl1997} subsequently
established that the competitive ratio of list scheduling is
$\Theta(\log m)$ and gave an $8$-competitive deterministic algorithm,
providing the first constant competitive guarantee for arbitrary
related machines. \citet{BermanCharikarKarpinski2000} improved the deterministic upper
bound to $3+\sqrt8\approx5.8284$ and obtained a randomized upper bound
of approximately $4.3111$. For two machines, \citet{EpsteinEtAl2001Two}
gave randomized algorithms with competitive ratio at most $1.5278$
for every speed ratio.

On the lower-bound side, \citet{EpsteinSgall2000} proved a bound of $2$
for randomized algorithms, while \citet{EbenlendrSgall2015} established
a lower bound of approximately $2.5650$ for deterministic algorithms. For small
numbers of machines, \citet{JezEtAl2013} proved deterministic lower bounds
of approximately $2.1414$ and $2.3146$ for four and five machines, respectively,
and obtained further bounds for up to eleven machines.

%% file: sections/02_deterministic_algorithm.tex
\section{The deterministic algorithm}
\label{sec:deterministic}

At each arrival, the algorithm first computes a lower bound $T_t$
on the optimal makespan $\OPT_t$ of $I_t$.
It then assigns the new job to the slowest machine whose load after
the assignment is at most $\alpha_0T_t$. Choosing the slowest feasible
machine uses slower capacity whenever the load threshold permits,
preserving fast-machine capacity for future large jobs.

\subsection{A lower bound on the offline optimum}
\label{sec:clock}

The average-work bound $\sum_{j\le t}p_j/\sum_{i=1}^m s_i$ and the
largest-job bound $\max_{j\le t}p_j/s_m$ do not fully account for the
capacity constraints imposed by large jobs. For example, consider
machines of speeds $1,2,8$ and two jobs of size $8$. These bounds
are $16/11$ and $1$, respectively. However, a schedule with makespan
less than $2$ would have to assign both jobs to the fastest machine,
where their total processing time is $2$.

To capture these capacity constraints, fix a candidate makespan
$y\ge0$ and a speed threshold $w\ge0$. Every job of size greater
than $wy$ must be assigned to a machine of speed greater than $w$.
Write
\[
S(w)=\sum_{i:s_i>w}s_i.
\]
Thus every schedule for $I_t$ with makespan at most $y$ satisfies
\begin{equation}
\label{eq:upper-tail-condition}
\sum_{\substack{j\le t\\p_j>wy}}p_j
\le yS(w)
\qquad\text{for every }w\ge0.
\end{equation}
These capacity constraints were used by \citet{BermanCharikarKarpinski2000}
to obtain lower bounds on the optimal makespan.

We set $T_0=0$ and define
\[
T_t=
\inf\left\{
y\ge0:
\sum_{\substack{j\le t\\p_j>wy}}p_j
\le yS(w)
\text{ for every }w\ge0
\right\}.
\]
Every $y<T_t$ is ruled out by at least one speed threshold, whereas
$\OPT_t$ satisfies all the constraints. Hence
$T_t\le\OPT_t$, and taking the infimum gives the strongest
lower bound supplied by these constraints. For $t\ge1$, the constraint
at $w=0$ ensures that every feasible $y$ is strictly positive.
The strict comparisons $p_j>wy$ and $s_i>w$ are intentional:
a job of size exactly $ys_i$ can be assigned to machine $i$
within makespan $y$.

List the job sizes in $I_t$ in non-increasing order as
$p_{(1)}\ge\cdots\ge p_{(t)}>0$, and define
\[
P_q=\sum_{k=1}^{q}p_{(k)},
\qquad
S_i=\sum_{k=i}^{m}s_k,
\]
with $P_0=0$. For a job of size $p$, define
\[
E_y(p)=\{i:p\le ys_i\}.
\]
These are precisely the machines on which the job can finish within
time $y$. Larger jobs can use fewer machines, so
\[
E_y(p_{(1)})
\subseteq
E_y(p_{(2)})
\subseteq\cdots\subseteq
E_y(p_{(t)}).
\]

In a fractional schedule, an amount $x_{ij}\ge0$ of job $j$ is
assigned to machine $i$, with $\sum_{i=1}^m x_{ij}=p_j$.
The load of machine $i$ is then $(\sum_{j\le t}x_{ij})/s_i$.

The next two lemmas explain how the capacity constraints yield a
computable lower bound. The first establishes the equivalence of
the capacity constraints, inequalities for the largest jobs, and
the existence of a fractional schedule satisfying the capacity
and assignment restrictions. The second gives a finite formula
for $T_t$, shows that the infimum is attained, and establishes
its basic properties.

\begin{lemma}
\label{lem:static}
For every $y>0$, the following three conditions are equivalent:
\begin{enumerate}[label=\textnormal{(\roman*)},leftmargin=*]
\item The capacity constraints in \eqref{eq:upper-tail-condition} hold.
\item For every $q\in\{1,\ldots,t\}$,
\begin{equation}
\label{eq:static-rank}
P_q\le y\sum_{i\in E_y(p_{(q)})}s_i.
\end{equation}
\item The input $I_t$ admits a fractional schedule
in which the total work of machine $i$ is at most $ys_i$, and
each job $j$ is assigned only to the machines in $E_y(p_j)$.
\end{enumerate}
\end{lemma}

\begin{proof}
\textbf{(i) $\Rightarrow$ (ii).}
Fix $q\in\{1,\ldots,t\}$. Let $w$ be the largest machine speed
outside $E_y(p_{(q)})$, or let $w=0$ if every machine belongs
to this set. Then $p_{(q)}>wy$ and
$\{i:s_i>w\}=E_y(p_{(q)})$. Every one of the $q$ largest jobs
therefore has size greater than $wy$. Applying \eqref{eq:upper-tail-condition} at this
speed threshold gives
\[
P_q
\le
\sum_{\substack{j\le t\\p_j>wy}}p_j
\le
yS(w)
=
y\sum_{i\in E_y(p_{(q)})}s_i,
\]
which proves (ii).

\textbf{(ii) $\Rightarrow$ (iii).}
Give machine $i$ a work capacity of $ys_i$. We construct a
fractional schedule by considering the jobs in non-increasing
order of size, breaking ties arbitrarily. Suppose that the first
$q-1$ jobs in this order have been assigned, and consider the
$q$-th largest job, whose size is $p_{(q)}$. For every $k<q$,
\[
E_y(p_{(k)})\subseteq E_y(p_{(q)}).
\]
Thus all previously assigned work, of total size $P_{q-1}$,
lies on the machines in $E_y(p_{(q)})$. By (ii), their total
unused capacity is
\[
y\sum_{i\in E_y(p_{(q)})}s_i-P_{q-1}
\ge
P_q-P_{q-1}
=
p_{(q)}.
\]
We can therefore assign the entire job to these machines without
exceeding their capacities. Repeating this step assigns every job
and produces a fractional schedule satisfying (iii).

\textbf{(iii) $\Rightarrow$ (i).}
Fix $w\ge0$. If $p_j>wy$, then every machine $i\in E_y(p_j)$
satisfies $ys_i\ge p_j>wy$, and hence $s_i>w$. All work belonging
to jobs of size greater than $wy$ must therefore be assigned to
machines of speed greater than $w$. Their total work capacity gives
\[
\sum_{\substack{j\le t\\p_j>wy}}p_j
\le
\sum_{i:s_i>w}ys_i
=
yS(w).
\]
Since $w$ was arbitrary, \eqref{eq:upper-tail-condition} holds.
\end{proof}

\begin{lemma}
\label{lem:clock-formula}
For every $t\ge1$, the infimum in the definition of $T_t$ is
attained, and
\begin{equation}
\label{eq:clock-formula}
T_t=
\max_{1\le q\le t}
\min_{1\le i\le m}
\max\left\{
\frac{p_{(q)}}{s_i},
\frac{P_q}{S_i}
\right\}.
\end{equation}
Moreover,
\begin{equation}
\label{eq:clock-mono}
0\le T_{t-1}\le T_t\le\OPT_t.
\end{equation}
\end{lemma}

\begin{proof}
Fix $q\in\{1,\ldots,t\}$ and $y>0$. We first show that \eqref{eq:static-rank}
for this $q$ holds if and only if there is a machine index $i$
such that
\[
p_{(q)}\le ys_i,
\qquad
P_q\le yS_i.
\]
Indeed, if \eqref{eq:static-rank} holds, then $E_y(p_{(q)})$ is nonempty because
$P_q>0$. Let $i$ be its smallest index. Since the machine speeds
are ordered, $E_y(p_{(q)})=\{i,\ldots,m\}$, and both inequalities
follow. Conversely, if the two inequalities hold for some $i$,
then every machine $i,\ldots,m$ belongs to $E_y(p_{(q)})$, and
\[
P_q\le yS_i
\le y\sum_{k\in E_y(p_{(q)})}s_k,
\]
which is \eqref{eq:static-rank}.

For a fixed $i$, the two inequalities are equivalent to
\[
y\ge
\max\left\{
\frac{p_{(q)}}{s_i},
\frac{P_q}{S_i}
\right\}.
\]
Consequently, every $y$ satisfying \eqref{eq:upper-tail-condition} is at least the right-hand
side of \eqref{eq:clock-formula}. Conversely, let $y$ equal the right-hand side of \eqref{eq:clock-formula},
which is positive. For each $q$, choose a machine index $i$
attaining the inner minimum, which exists because there are
finitely many machines. Then
\[
\max\left\{
\frac{p_{(q)}}{s_i},
\frac{P_q}{S_i}
\right\}\le y.
\]
By the equivalence established above, \eqref{eq:static-rank} holds for every $q$.
Lemma~\ref{lem:static} therefore implies that this value of $y$ satisfies \eqref{eq:upper-tail-condition}.
Thus it equals $T_t$, and the infimum is attained.

Adding a job can only increase the left-hand side of \eqref{eq:upper-tail-condition}, so
$T_t$ is non-decreasing. Together with $T_0=0$, this gives
$0\le T_{t-1}\le T_t$. Finally, every optimal schedule for $I_t$
satisfies \eqref{eq:upper-tail-condition} with $y=\OPT_t$, giving
$T_t\le\OPT_t$.
\end{proof}

Thus $T_t$ is the smallest candidate makespan satisfying all the
capacity constraints. Formula \eqref{eq:clock-formula} separates two requirements.
The term $p_{(q)}/s_i$ checks whether the $q$-th largest job can
use machine $i$, while $P_q/S_i$ measures the total work of the
$q$ largest jobs relative to the combined speed of machines
$i,\ldots,m$. The minimum chooses the best machine index for the
constraint corresponding to $q$, and the maximum enforces all
these constraints.

Two useful consequences follow by applying \eqref{eq:upper-tail-condition} at $y=T_t$,
first with $w=s_m$ and then with $w=0$:
\[
p_j\le T_t s_m
\quad (j\le t),
\qquad
\sum_{j\le t}p_j\le T_tS_1.
\]

\subsection{The assignment rule and its implementation}
\label{sec:algorithm}

After computing $T_t$, the algorithm assigns job $t$ to the
smallest machine index $i$ satisfying
\begin{equation}
\label{eq:first-fit}
\frac{W_i(t-1)+p_t}{s_i}\le\alpha_0T_t.
\end{equation}
The left-hand side is the load of machine $i$ after receiving
the new job. The lower bound $T_t$ is computed from $I_t$,
including the new job, while $W_i(t-1)$ includes all work
previously assigned to machine $i$.

\begin{algorithm}[htbp]
\caption{First fit with the lower bound $T_t$}
\label{alg:main}
\begin{enumerate}[label=\arabic*.,leftmargin=*,itemsep=3pt]
\item Order the machines by non-decreasing speed, breaking ties
by fixed identifiers. Initialize $W_i=0$ for every machine $i$
and $T_0=0$.
\item Upon the arrival of job $t$, compute $T_t$ from the job
sizes in $I_t$ using \eqref{eq:clock-formula}.
\item Choose the smallest index $i$ such that
\[
\frac{W_i+p_t}{s_i}\le\alpha_0T_t.
\]
Assign job $t$ to machine $i$, update $W_i\leftarrow W_i+p_t$,
and continue with the next arrival.
\end{enumerate}
\end{algorithm}

The preference for the slowest feasible machine is also used in
the 8-competitive algorithm of \citet{AspnesEtAl1997}. \citet{BermanCharikarKarpinski2000} use the capacity constraints in \eqref{eq:upper-tail-condition} to check their estimate
after each arrival and multiply it by a fixed factor whenever a
constraint is violated. Their assignment limits depend on both
the current estimate and the estimates used in earlier phases.
Our algorithm computes the smallest feasible value $T_t$ after
every arrival and applies the threshold $\alpha_0T_t$ directly
to the resulting machine load, using the total work accumulated
over all earlier arrivals.

For a short example, consider machines of speeds $(1,2,8)$ and
jobs of sizes $(1,1,2,1)$. Table~\ref{tab:upper-short-run} shows the assignments made by Algorithm~\ref{alg:main}.
The last arrival illustrates that $T_t$ can remain unchanged
even when the total work increases.

\begin{table}[htbp]
\centering
\small
\caption{An example run of Algorithm~\ref{alg:main}. The last column gives the total work of each machine.}
\label{tab:upper-short-run}
\begin{tabular}{@{}ccccc@{}}
\toprule
$t$ & $p_t$ & $T_t$ & Assigned machine
& $(W_1(t),W_2(t),W_3(t))$ \\
\midrule
$1$ & $1$ & $1/8$ & $2$ & $(0,1,0)$ \\
$2$ & $1$ & $1/4$ & $1$ & $(1,1,0)$ \\
$3$ & $2$ & $1/2$ & $2$ & $(1,3,0)$ \\
$4$ & $1$ & $1/2$ & $1$ & $(2,3,0)$ \\
\bottomrule
\end{tabular}
\end{table}

\paragraph{Time and space complexity.}
Sorting the machines initially takes $O(m\log m)$ comparisons.
The sums $S_i$ can then be computed in $O(m)$ operations by one
backward pass and reused at every arrival. Maintain the job sizes
in a sorted array. At arrival $t$, inserting $p_t$ into the existing
array takes $O(t)$ operations, and computing the updated sums
$P_1,\ldots,P_t$ takes another $O(t)$ operations.

To evaluate \eqref{eq:clock-formula}, fix $q$. As $i$ increases, $p_{(q)}/s_i$ is
non-increasing, whereas $P_q/S_i$ is increasing. The minimum of
their maximum is therefore attained at the first index where
$p_{(q)}/s_i\le P_q/S_i$, or at its predecessor if one exists.
This first index exists because the inequality holds at $i=m$.
The comparison can be written equivalently as
$S_i/s_i\le P_q/p_{(q)}$. Since $S_i/s_i$ decreases with $i$
and $P_q/p_{(q)}$ increases with $q$, the first such index moves
only to the left as $q$ increases. During a single arrival, a
shared pointer therefore moves at most $m-1$ times while processing
all $q=1,\ldots,t$. Each $q$ requires only constant additional work
to evaluate the candidate indices and update the maximum in \eqref{eq:clock-formula},
giving $O(t+m)$ operations for this evaluation.

Once $T_t$ is known, scanning the machines to find the first one
satisfying \eqref{eq:first-fit} takes $O(m)$ operations, and updating its total work
takes constant time. Combining insertion, computation of the sums
$P_q$, evaluation of \eqref{eq:clock-formula}, and machine selection gives $O(t+m)$
operations at arrival $t$. Summing over all $n$ arrivals and
including the initial sorting and precomputation gives
\[
O(m\log m)+O(m)+\sum_{t=1}^{n}O(t+m)
=
O(n^2+mn+m\log m).
\]
At arrival $t$, the stored data consist of the $t$ job sizes and
their cumulative sums, together with the machine speeds, the sums $S_i$,
and the current work totals $W_i$. This requires $O(t+m)$ space,
so the space requirement for the entire input is $O(n+m)$.

%% file: sections/03_upper_bound.tex
\section{Analysis of the deterministic algorithm}
\label{sec:upper-analysis}

\begin{theorem}
\label{thm:analytic}
Algorithm~\ref{alg:main} is $\alpha_0$-competitive for any number of related machines.
\end{theorem}

The proof has two main steps. We first combine the fractional schedules
for $I_t$, $1\le t\le n$, into a fixed fractional schedule with suitable
load bounds after every arrival. We then show that Algorithm~\ref{alg:main} assigns
at least as much total work to machines $1,\ldots,h$ as this fractional
schedule does, for every $h$. This comparison ensures that the assignment
rule always finds a feasible machine, yielding
$C_{\max}(t)\le\alpha_0T_t\le\alpha_0\OPT_t$.

\subsection{A fixed fractional schedule feasible after every arrival}
\label{subsec:fixed-fractional-allocation}

Lemma~\ref{lem:clock-formula} shows that $T_t$ satisfies the capacity constraints.
Lemma~\ref{lem:static} then provides a fractional schedule for $I_t$ at makespan $T_t$.
These schedules may assign an earlier job to different machines as more
jobs arrive. To compare them with the online algorithm after every
arrival, we construct a single fractional schedule whose amounts assigned
from each job to each machine remain fixed.

For example, consider machines with speeds $(1,3/2)$ and two jobs of
sizes $3$ and $9/2$, arriving in this order. The largest-job bounds give
$T_1\ge 2$ and $T_2\ge 3$. These bounds are attained by assigning the first
job to the fast machine and by assigning the two jobs to the slow and fast
machines, respectively. Thus $T_1=2$ and $T_2=3$. Under the eligibility
restrictions of Lemma~\ref{lem:static}, the first job must use the fast machine at $T_1$.
At $T_2$, the second job must use the fast machine and fills its entire
capacity, so the first job must use the slow machine. The two fractional
schedules therefore require different allocations of the first job.

Fix an input $I=(p_1,\ldots,p_n)$. Let $f_{ij}\ge 0$ denote the amount of
job $j$ assigned to machine $i$ in a fractional schedule. Thus $f_{ij}/p_j$
is the fraction of the job assigned to that machine. Define
\[
F_i(t)=\sum_{j=1}^{t}f_{ij},
\qquad F_i(0)=0.
\]
Here $F_i(t)$ is the total work of machine $i$ contributed by the jobs in
$I_t$. We choose the amounts $f_{ij}$ once for the entire input, so
$F_i(t)$ changes only by adding the amount assigned to the newly arrived
job. This fractional schedule is used in the analysis of Algorithm~\ref{alg:main}.

We require every job to be fully assigned, and we seek bounds both on the
total work of each machine after every arrival and on the processing time
of a whole job on any machine receiving part of it. For prescribed bounds,
these requirements form a finite system of linear constraints in the
variables $f_{ij}$. The following lemma satisfies them by an explicit
construction.

\begin{lemma}
\label{lem:comparator}
Let $r>1$ and $c=r/\log r$. There are nonnegative amounts $f_{ij}$, for
$1\le i\le m$ and $1\le j\le n$, satisfying
\begin{equation}
\label{eq:comparator-support}
\sum_{i=1}^{m}f_{ij}=p_j,
\qquad
f_{ij}>0\implies p_j\le rT_js_i,
\end{equation}
and
\begin{equation}
\label{eq:comparator-history}
F_i(t)\le cT_ts_i
\qquad
(1\le i\le m,\ 0\le t\le n).
\end{equation}
\end{lemma}

\begin{proof}
By Lemma~\ref{lem:clock-formula}, the capacity constraints in \eqref{eq:upper-tail-condition} hold at $T_q$ for every
$1\le q\le n$. Taking $w=s_m$ gives $T_q\ge p_q/s_m>0$, so Lemma~\ref{lem:static}
provides a fractional schedule for $I_q$ in which machine $i$ receives at
most $T_qs_i$ work and job $j$ uses only machines in $E_{T_q}(p_j)$.
Choose one such schedule for each $q$, and let $x_{ij}^{(q)}$ be the amount
of job $j$ assigned to machine $i$. Set $x_{ij}^{(q)}=0$ for $j>q$, and let
$x_{ij}^{(0)}=0$.

For $y>0$, define
\[
q(y)=\max\{0\le q\le n:T_q\le y\},
\qquad
x_{ij}(y)=x_{ij}^{(q(y))}.
\]
Since $T_q$ is non-decreasing, this schedule includes exactly the jobs with
$T_j\le y$. Taking the largest index includes every job sharing the same
value of $T_j$. The functions $x_{ij}(y)$ are constant between consecutive
distinct values of $T_q$, and $q(y)=n$ for $y\ge T_n$. For every $y>0$, the
chosen schedule satisfies
\begin{equation}
\label{eq:comparison-static-properties}
\sum_{j=1}^{n}x_{ij}(y)\le ys_i,
\qquad
x_{ij}(y)>0\implies p_j\le ys_i.
\end{equation}
Moreover, $\sum_{i=1}^{m}x_{ij}(y)=p_j$ whenever $y\ge T_j$.

For each job $j$, we define
\[
f_{ij}
=
\frac1{\log r}
\int_{T_j}^{rT_j}\frac{x_{ij}(y)}{y}\,dy.
\]
Every schedule used in this integral includes job $j$, and any machine
receiving part of it can process the whole job within time $rT_j$.
Dividing by $y$ gives the same capacity bound $s_i$ for machine $i$ at
every $y$. The factor $1/\log r$ preserves the total size of each job,
since $\int_{T_j}^{rT_j}dy/y=\log r$.

Indeed,
\[
\begin{aligned}
\sum_{i=1}^{m}f_{ij}
&=
\frac{1}{\log r}
\int_{T_j}^{rT_j}
\sum_{i=1}^{m}x_{ij}(y)\,\frac{dy}{y}\\
&=
\frac{p_j}{\log r}
\int_{T_j}^{rT_j}\frac{dy}{y}
=p_j.
\end{aligned}
\]
If $p_j>rT_js_i$, then $p_j>ys_i$ throughout the interval $[T_j,rT_j]$.
By \eqref{eq:comparison-static-properties}, $x_{ij}(y)=0$ throughout this interval, so $f_{ij}=0$.
This proves both assertions in \eqref{eq:comparator-support}.

For $t=0$, \eqref{eq:comparator-history} follows from $F_i(0)=T_0=0$. Fix a machine $i$ and
$1\le t\le n$. For every $j\le t$, monotonicity gives
$[T_j,rT_j]\subseteq(0,rT_t]$. We can therefore write
\[
\begin{aligned}
(\log r)F_i(t)
&=
\int_0^{rT_t}
\sum_{j=1}^{t}
\mathbf{1}_{\{T_j\le y\le rT_j\}}
x_{ij}(y)\,\frac{dy}{y}\\
&\le
\int_0^{rT_t}
\sum_{j=1}^{n}x_{ij}(y)\,\frac{dy}{y}\\
&\le
\int_0^{rT_t}ys_i\,\frac{dy}{y}
=rT_ts_i.
\end{aligned}
\]
At each $y$, the first integrand selects some of the nonnegative amounts
in the single schedule $x(y)$. Their sum is therefore at most the total
work of machine $i$ in that schedule, even when the schedule includes jobs
arriving after $t$. The second inequality follows from \eqref{eq:comparison-static-properties}. Dividing by
$\log r$ proves \eqref{eq:comparator-history} for every $t$, using the same amounts $f_{ij}$.
\end{proof}

The integral construction can also be written as a finite weighted
average. For job $j$, subdivide $[T_j,rT_j]$ at the distinct values of $T_q$
in its interior, obtaining endpoints
\[
u_0=T_j<u_1<\cdots<u_L=rT_j.
\]
Since $x(y)=x(u_a)$ on $[u_a,u_{a+1})$,
\[
f_{ij}
=
\sum_{a=0}^{L-1}
\frac{\log(u_{a+1}/u_a)}{\log r}\,
x_{ij}(u_a).
\]
The weights are nonnegative and sum to one, since
\[
\sum_{a=0}^{L-1}\log(u_{a+1}/u_a)
=
\log(u_L/u_0)
=
\log r.
\]

Returning to the two-machine example, take $r=2$. Choose $x(y)$ to assign
the first job to the fast machine for $2\le y<3$, and to assign the two
jobs to the slow and fast machines, respectively, for $y\ge3$. Job 1 is
averaged over $[2,4]$, giving
\[
f_{11}
=
3\frac{\log(4/3)}{\log2},
\qquad
f_{21}
=
3\frac{\log(3/2)}{\log2}.
\]
These amounts sum to $3$. Job 2 is averaged over $[3,6]$, throughout which
it is assigned to the fast machine, so $f_{12}=0$ and $f_{22}=9/2$. The
same amounts $f_{11}$ and $f_{21}$ contribute to the machine totals both
before and after job 2 arrives.

\subsection{Feasibility of the assignment rule}
\label{subsec:first-fit-feasibility}

We compare the total work assigned by the online and fractional schedules
to machines $1,\ldots,h$. Fix an input $I=(p_1,\ldots,p_n)$, let $r>1$,
and put $c=r/\log r$. Choose a fractional schedule for this input as
provided by Lemma~\ref{lem:comparator}, with amounts $f_{ij}$ and cumulative work $F_i(t)$.
Let $W_i(t)$ denote the total work of machine $i$ under Algorithm~\ref{alg:main} after
the first $t$ arrivals. For $1\le h\le m$, define
\[
A_h(t)=\sum_{i=1}^{h}W_i(t),
\qquad
D_h(t)=\sum_{i=1}^{h}F_i(t),
\]
and set $A_0(t)=D_0(t)=0$. Thus $A_h(t)$ and $D_h(t)$ are the total work
assigned to the $h$ slowest machines by the online and fractional
schedules, respectively. We will show that
\begin{equation}
\label{eq:slow-machine-dom}
A_h(t)\ge D_h(t)
\qquad (0\le h\le m)
\end{equation}
after every arrival, and use this inequality to rule out a failed
assignment.

\begin{lemma}
\label{thm:parametric}
If $r+r/\log r\le\alpha_0$, then Algorithm~\ref{alg:main} assigns every
job and satisfies $C_{\max}(t)\le\alpha_0T_t$ after every arrival.
\end{lemma}

\begin{proof}
We prove inductively that every assignment succeeds and that \eqref{eq:slow-machine-dom} holds
for every $h$.
At $t=0$, both sides of \eqref{eq:slow-machine-dom} are zero.

Consider job $t$, assuming the claim after the first
$t-1$ arrivals. In the fractional schedule, this job is assigned only
to the machines in $E_{rT_t}(p_t)=\{k,\ldots,m\}$, where
\[
k=\min\{i:p_t\le rT_ts_i\}.
\]
This index exists because $p_t\le T_ts_m$ and $r>1$. Let $b$ be the first
machine satisfying the assignment condition \eqref{eq:first-fit}. If no machine satisfies
it, set $b=m+1$ and, for this hypothetical failed assignment, put
$W_i(t)=W_i(t-1)$ for every $i$. The quantities $F_i(t)$ include the
amounts $f_{it}$ of the new job in either case. Hence
\[
\begin{aligned}
A_h(t)-A_h(t-1)&=p_t\,\mathbf{1}_{\{h\ge b\}},\\
D_h(t)-D_h(t-1)&=\sum_{i=1}^{h}f_{it}.
\end{aligned}
\]
The second increment lies between $0$ and $p_t$, and is zero when $h<k$.
The index $k$ specifies which machines may receive a fractional share of job $t$.
Algorithm~\ref{alg:main} may accept the job at a machine with index $b<k$.

If $h\ge b$, the online schedule adds the whole job to machines
$1,\ldots,h$, whereas their total work in the fractional schedule
increases by at most $p_t$. The induction hypothesis gives
\[
A_h(t)=A_h(t-1)+p_t
\ge D_h(t-1)+p_t
\ge D_h(t).
\]

If $h<b$ and $h<k$, neither schedule adds any work to machines
$1,\ldots,h$, so \eqref{eq:slow-machine-dom} follows directly from the induction hypothesis.

If $k\le h<b$, the online schedule adds no work to machines
$1,\ldots,h$, although the fractional schedule may add part of the job
there. Every machine $i=k,\ldots,h$ was considered and rejected the job.
Since the assignment condition accepts equality,
$W_i(t-1)+p_t>\alpha_0T_ts_i$. Also $p_t\le rT_ts_i$ because $i\ge k$.
Consequently,
\begin{equation}
\label{eq:rejected-machine}
W_i(t)=W_i(t-1)>\alpha_0T_ts_i-p_t\ge(\alpha_0-r)T_ts_i\ge cT_ts_i\ge F_i(t).
\end{equation}
Thus the work already assigned to each of these machines by the online
algorithm exceeds its total work in the fractional schedule, including
the amount assigned from the new job. On machines $1,\ldots,k-1$, neither
schedule adds work, so
\[
A_{k-1}(t)=A_{k-1}(t-1)
\ge D_{k-1}(t-1)=D_{k-1}(t).
\]
Combining these inequalities yields
\[
A_h(t)-D_h(t)=A_{k-1}(t)-D_{k-1}(t)+\sum_{i=k}^{h}\bigl(W_i(t)-F_i(t)\bigr)\ge0.
\]
This also covers $k=1$, since $A_0(t)=D_0(t)=0$.

The three cases establish \eqref{eq:slow-machine-dom}, including when $b=m+1$. If all machines
rejected the job, however, the online schedule would contain only the
first $t-1$ jobs, while the fractional schedule would include the new job.
Since $p_t>0$, this would give
\[
A_m(t)=\sum_{j=1}^{t-1}p_j
<
\sum_{j=1}^{t}p_j
=D_m(t),
\]
contradicting \eqref{eq:slow-machine-dom}. Every assignment therefore succeeds.

It remains to bound the machine loads. Initially every machine has zero
work. At arrival $t$, the chosen machine satisfies
$W_b(t)\le\alpha_0T_ts_b$ by \eqref{eq:first-fit}. For every other machine, induction and
the monotonicity of $T_t$ give
\[
W_i(t)=W_i(t-1)
\le\alpha_0T_{t-1}s_i
\le\alpha_0T_ts_i.
\]
Dividing by the machine speeds and taking the maximum proves
$C_{\max}(t)\le\alpha_0T_t$ for every $1\le t\le n$.
\end{proof}

\subsection{The competitive guarantee}
\label{subsec:threshold-guarantee}

The parameters $c$ and $r$ control two different quantities. In the
fractional schedule, the load of machine $i$ after the first $t$ arrivals
is at most $cT_t$. If this machine receives part of job $t$, it can process
the whole job within time $rT_t$. The condition $c+r\le\alpha_0$ therefore
allows \eqref{eq:rejected-machine} to compare the work already assigned by the online algorithm
with the updated fractional work. Since $c=r/\log r$, we choose $r$ to
minimize $r+r/\log r$.

\begin{proof}[Proof of Theorem~\ref{thm:analytic}]
For $r>1$, let $g(r)=r+r/\log r$. Its derivative is
\[
g'(r)=\frac{(\log r)^2+\log r-1}{(\log r)^2}.
\]
The numerator is negative for $0<\log r<\theta$ and positive for
$\log r>\theta$, where $\theta=(\sqrt5-1)/2$. Hence the unique minimum is
attained at $r^*=e^\theta$, with
\[
g(r^*)
=
\frac{3+\sqrt5}{2}
\exp\!\left(\frac{\sqrt5-1}{2}\right)
=\alpha_0.
\]
Applying Lemma~\ref{thm:parametric} with $r=r^*$ and using \eqref{eq:clock-mono}, we obtain
\[
C_{\max}(t)\le\alpha_0T_t\le\alpha_0\OPT_t
\]
after every arrival. This proves that Algorithm~\ref{alg:main} is
$\alpha_0$-competitive.
\end{proof}

The competitive ratio of Algorithm~\ref{alg:main} is exactly $\alpha_0$. Consider two
machines with speeds $(1,\alpha_0)$ and a single job of size $1$. Assigning
the job to the fast machine gives $\OPT_1=1/\alpha_0$. The
largest-job bound and $T_1\le\OPT_1$ imply $T_1=1/\alpha_0$.
The slow machine satisfies the assignment condition with equality,
since $\alpha_0T_1s_1=1$. Algorithm~\ref{alg:main} therefore assigns the job to that
machine, obtaining makespan $1$ and ratio $\alpha_0$.

%% file: sections/04_lower_bound.tex
\section{A randomized lower bound of \texorpdfstring{$e$}{e}}
\label{sec:lower-bound}

\begin{theorem}
\label{thm:lb-main}
For every $0<\kappa<e$ and all sufficiently large $m$, there exists
a speed vector $\boldsymbol{s}\in\mathbb{Q}_{>0}^{m}$ such that,
for every randomized online algorithm $\mathcal A$, there is a
finite input $I$ with positive rational job sizes satisfying
\[
   \mathbb{E}\!\left[C_{\max}^{\mathcal A}(I)\right]
   >\kappa\,\OPT(I).
\]
Consequently, the competitive ratio of every randomized online algorithm is at least $e$.
\end{theorem}

We construct machines in geometric speed levels and a fixed sequence
of job batches. We first give offline schedules for the jobs
revealed through each batch. We then bound the work assigned by an
online algorithm to each speed level using its makespans at selected
batch boundaries. For suitable parameters, taking expectations and
comparing these bounds with the total work of the input shows that,
after some fixed number of arrivals, the expected makespan exceeds
$\kappa$ times the offline optimum.

\subsection{The construction}
\label{subsec:lower-construction}

Choose an integer $H\ge3$ and set $K=H^2$. For each level $\ell=0,\ldots,K-1$, take
$2^\ell$ machines of speed $2^{-\ell}$. Each level therefore has total speed one, and
the total number of machines is
\[
m_0=\sum_{\ell=0}^{K-1}2^\ell=2^K-1.
\]
Table~\ref{tab:lower-machines} lists the levels from fastest to slowest. Geometric speed
levels and increasing job sizes also appear in earlier lower bounds for related machines
\citep{EpsteinSgall2000,EbenlendrSgall2015}.

\begin{table}[htbp]
\centering
\small
\caption{The machine levels in the lower-bound construction.}
\label{tab:lower-machines}
\begin{tabular}{cccc}
\toprule
Level & Number of machines & Speed of each machine & Total speed\\
\midrule
$0$ & $1$ & $1$ & $1$\\
$1$ & $2$ & $1/2$ & $1$\\
$2$ & $4$ & $1/4$ & $1$\\
$\vdots$ & $\vdots$ & $\vdots$ & $\vdots$\\
$K-1$ & $2^{K-1}$ & $2^{-(K-1)}$ & $1$\\
\bottomrule
\end{tabular}
\end{table}

For each $H\le i<K$, batch $B_i$ consists of $a_i$ jobs of size $2^{-i}$, where
\begin{equation}
\label{eq:lower-batches}
a_i=\left\lfloor\frac{e(H-2)2^i}{i+1}\right\rfloor.
\end{equation}
Reveal the batches in the order
\[
B_{K-1},B_{K-2},\ldots,B_H,
\]
as listed in Table~\ref{tab:lower-batches}. Thus the jobs arrive in non-decreasing order
of size. Since $H\ge3$ and $2^i/(i+1)\ge2$ for $i\ge3$, every batch is nonempty. All
machine speeds and job sizes are positive rationals.

\begin{table}[htbp]
\centering
\small
\caption{The batches in arrival order. The counts $a_i$ are defined
in \eqref{eq:lower-batches}.}
\label{tab:lower-batches}
\begin{tabular}{cccc}
\toprule
Arrival order & Batch & Job size & Number of jobs\\
\midrule
First & $B_{K-1}$ & $2^{-(K-1)}$ & $a_{K-1}$\\
Second & $B_{K-2}$ & $2^{-(K-2)}$ & $a_{K-2}$\\
$\vdots$ & $\vdots$ & $\vdots$ & $\vdots$\\
Last & $B_H$ & $2^{-H}$ & $a_H$\\
\bottomrule
\end{tabular}
\end{table}

Let $I=(p_1,\ldots,p_n)$ denote this fixed job sequence. Write $h_i$ for the arrival
index of the last job in batch $B_i$, and $W$ for the total work of the sequence. Then
\[
h_i=\sum_{r=i}^{K-1}a_r,
\qquad
n=h_H,
\qquad
W=\sum_{i=H}^{K-1}a_i2^{-i}.
\]

The job counts in \eqref{eq:lower-batches} are chosen to ensure $\OPT_h\le\tau_i$ after
every arrival in batch $B_i$, where $\tau_i=H/i\le1$. To explain the coefficient $e$,
fix $H\le i\le j<K$. Among the jobs in $I_{h_i}$, those of size at least $2^{-j}$ are
precisely the jobs in batches $B_i,\ldots,B_j$. Such jobs cannot be processed on levels
greater than $j$ within time $\tau_i$. Levels $0,\ldots,j$ have total speed $j+1$,
giving total work capacity $(j+1)H/i$ by that time. If each batch $B_r$ has total work
at most $\eta H/(r+1)$, for a coefficient $\eta>0$, then the total work of these batches
is at most
\[
\eta H\sum_{r=i}^{j}\frac1{r+1}
\le
\eta H\log\frac{j+1}{i}.
\]
Writing $x=(j+1)/i$, this upper bound fits within the capacity whenever $\eta\log x\le
x$. The largest coefficient satisfying this inequality for every $x>1$ is $e$.
Subtracting $2$ from $H$ provides slack for the loss from rounding machine capacities
down to whole-job slots. Lemma~\ref{lem:lower-integral-schedules} gives the precise
rounding argument and constructs offline schedules establishing the desired bound.

\subsection{Offline schedules}
\label{subsec:lower-offline}

\begin{lemma}
\label{lem:lower-integral-schedules}
If job $h$ belongs to batch $B_i$, then $I_h$ admits an offline schedule of makespan at
most $\tau_i=H/i$. In particular, $0<\OPT_h\le1$ for $1\le h\le n$.
\end{lemma}

\begin{proof}
Fix $H\le i<K$ and set $\tau=\tau_i\le1$. We first construct a schedule for $I_{h_i}$,
consisting of batches $B_i,\ldots,B_{K-1}$, with makespan at most $\tau$.

For $i\le j<K$, set $u=2^{-j}$. A machine at level $\ell$ has work capacity
$\tau2^{-\ell}$ within time $\tau$, so it can accommodate $\lfloor\tau2^{j-\ell}\rfloor$
whole jobs of size $u$. We call each such place a slot. Levels $\ell>j$ provide no slots
because $\tau\le1$. The total work that these slots can hold is
\[
\Psi_j(\tau)
=
2^{-j}\sum_{\ell=0}^{j}
2^\ell\left\lfloor\tau2^{j-\ell}\right\rfloor.
\]
Using $\lfloor y\rfloor>y-1$, we obtain
\[
\begin{aligned}
\Psi_j(\tau)
&>
2^{-j}\sum_{\ell=0}^{j}
2^\ell\bigl(\tau2^{j-\ell}-1\bigr)\\
&=
(j+1)\tau-2^{-j}\sum_{\ell=0}^{j}2^\ell\\
&>
(j+1)\tau-2.
\end{aligned}
\]
Thus the total work capacity lost through rounding is less than $2$.

On the other hand, \eqref{eq:lower-batches} and $e\log x\le x$ give
\begin{equation}
\label{eq:lower-work-slots}
\begin{aligned}
\sum_{r=i}^{j}a_r2^{-r}
&\le
e(H-2)\sum_{r=i}^{j}\frac1{r+1}
\le
e(H-2)\log\frac{j+1}{i}\\
&\le
(H-2)\frac{j+1}{i}
=
(j+1)\tau-\frac{2(j+1)}{i}\\
&\le
(j+1)\tau-2
<
\Psi_j(\tau).
\end{aligned}
\end{equation}
The harmonic-sum bound follows by integrating $1/x$ over each interval $[r,r+1]$.

We now assign batches $B_i,B_{i+1},\ldots,B_{K-1}$ in decreasing order of job size.
Suppose that $B_i,\ldots,B_{j-1}$ have been assigned within makespan $\tau$; for $j=i$,
the schedule is empty. Every job in an earlier batch $B_r$, with $i\le r<j$, has size
$2^{-r}=2^{j-r}u$. Since $2^{j-r}$ is an integer, assigning such a job reduces the
number of available slots for jobs of size $u$ on its machine by exactly $2^{j-r}$. Thus
the previously assigned batches occupy a total of $\sum_{r=i}^{j-1}a_r2^{j-r}$ slots,
regardless of their distribution among the machines.

The initial total number of slots is $\Psi_j(\tau)/u$. After subtracting the slots
occupied by the previously assigned batches, the remaining number is
\[
\frac{\Psi_j(\tau)}{u}
-\sum_{r=i}^{j-1}a_r2^{j-r}
=
\frac{\displaystyle
\Psi_j(\tau)-\sum_{r=i}^{j-1}a_r2^{-r}}{u}
\ge
\frac{a_j2^{-j}}{u}
=
a_j,
\]
where the inequality follows from \eqref{eq:lower-work-slots}. The existing schedule has
makespan at most $\tau$, so each machine has a nonnegative number of remaining slots. We
can therefore assign all $a_j$ jobs of $B_j$, one at a time, to machines with remaining
slots. This preserves the makespan bound $\tau$ and completes the induction.

The resulting schedule contains all jobs in $I_{h_i}$. If job $h$ belongs to $B_i$,
deleting jobs $h+1,\ldots,h_i$ gives a schedule for $I_h$ with makespan at most
$\tau_i\le1$. Therefore $\OPT_h\le\tau_i$. Since every job has positive size, $\OPT_h>0$
for every $1\le h\le n$.
\end{proof}

\subsection{Analysis of online algorithms}
\label{subsec:lower-fixed-length}

Fix $R\ge1$. We separate assigned jobs according to whether their processing time on the
chosen machine is at most $R$ or exceeds $R$. We will bound the work of the first group
using the makespans at fixed arrival indices, and the work of the second group using the
final makespan.

Put $d=\lfloor\log_2R\rfloor$. A job in batch $B_i$ has processing time $2^{\ell-i}$ on
a machine of level $\ell$. Since $\ell-i$ is an integer, $2^{\ell-i}\le R$ holds if and
only if $i\ge\ell-d$. The batches arrive in decreasing order of their indices. Define
\[
r_\ell=h_{\max\{H,\ell-d\}}
\qquad(0\le\ell<K).
\]
If $\ell-d\le H$, then $r_\ell=n$. Otherwise, $r_\ell$ is the arrival index of the last
job in $B_{\ell-d}$. By arrival $r_\ell$, every job whose processing time on a machine
of level $\ell$ is at most $R$ has arrived, and every later job takes more than $R$ time
units on such a machine.

Extend the offline bounds $\tau_j$ to all integers $j<K$ by setting
\[
\tau_j=
\begin{cases}
1, & j<H,\\
H/j, & H\le j<K.
\end{cases}
\]
Lemma~\ref{lem:lower-integral-schedules} gives
\begin{equation}
\label{eq:lower-cutoff-opt}
\OPT_{r_\ell}\le\tau_{\ell-d}.
\end{equation}
We will sum the work bounds over all speed levels, so define
\[
D_R=\sum_{\ell=0}^{K-1}\tau_{\ell-d}.
\]

\begin{lemma}
\label{lem:lower-fixed-length}
Run any randomized online algorithm $\mathcal A$ on the fixed sequence $I$. Then
\begin{equation}
\label{eq:lower-fixed-length-ratio}
\rho:=
\max_{1\le h\le n}
\frac{\mathbb{E}[C_{\max}(h)]}{\OPT_h}
\ge
\frac{W}{D_R+W/R}.
\end{equation}
\end{lemma}

\begin{proof}
Fix a realization of the algorithm's random choices. Let $G_\ell$ be the total work of
jobs assigned to level $\ell$ whose processing time on that level is at most $R$. Let
$V$ be the total work of all jobs whose processing time on their assigned machines
exceeds $R$. Every job belongs to exactly one of these groups, so
\[
W=\sum_{\ell=0}^{K-1}G_\ell+V.
\]

Every job counted in $G_\ell$ arrives by $r_\ell$. Therefore,
\begin{equation}
\label{eq:lower-good-work}
G_\ell
\le
\sum_{a\text{ on level }\ell}W_a(r_\ell)
\le
C_{\max}(r_\ell)
\sum_{a\text{ on level }\ell}s_a
=
C_{\max}(r_\ell).
\end{equation}
Here the second inequality follows from $W_a(r_\ell)/s_a\le C_{\max}(r_\ell)$, and the
equality uses the fact that each level has total speed one.

If $V>0$, some job has processing time greater than $R$ on its assigned machine, so
$C_{\max}(n)>R$. Together with $V\le W$, this implies
\begin{equation}
\label{eq:lower-bad-work}
V\le\frac{W}{R}C_{\max}(n).
\end{equation}
The same inequality holds when $V=0$. Combining \eqref{eq:lower-good-work} and
\eqref{eq:lower-bad-work}, we obtain
\begin{equation}
\label{eq:lower-pathwise-work}
W
\le
\sum_{\ell=0}^{K-1}C_{\max}(r_\ell)
+
\frac{W}{R}C_{\max}(n)
\end{equation}
for every realization. Several levels may have the same index $r_\ell$; their
contributions remain separate because they account for jobs assigned to different
levels.

Taking expectations in \eqref{eq:lower-pathwise-work} and using the definition of $\rho$
gives
\[
\begin{aligned}
W
&\le
\sum_{\ell=0}^{K-1}\mathbb{E}[C_{\max}(r_\ell)]
+
\frac{W}{R}\mathbb{E}[C_{\max}(n)]\\
&\le
\rho\left(
\sum_{\ell=0}^{K-1}\OPT_{r_\ell}
+
\frac{W}{R}\OPT_n
\right)\\
&\le
\rho(D_R+W/R).
\end{aligned}
\]
The last inequality uses \eqref{eq:lower-cutoff-opt} and $\OPT_n\le1$. All ratios
defining $\rho$ are finite: Lemma~\ref{lem:lower-integral-schedules} gives $\OPT_h>0$,
and every realized makespan is at most $W/\min_a s_a$. Dividing by $D_R+W/R>0$ proves
\eqref{eq:lower-fixed-length-ratio}.
\end{proof}

For each algorithm, an index $h$ attaining the maximum in
\eqref{eq:lower-fixed-length-ratio} can be fixed before its random choices are realized.
The index may depend on the algorithm, but not on the outcomes of those random choices.

\subsection{The asymptotic lower bound}
\label{subsec:lower-limit}

We now choose $R$ and $H$ so that the bound in Lemma~\ref{lem:lower-fixed-length}
exceeds any prescribed $\kappa<e$. We first estimate $W/D_R$ as $H$ tends to infinity
with $R$ fixed.

\begin{lemma}
\label{lem:lower-harmonic}
For every fixed $R\ge1$, the following limit holds as $H\to\infty$:
\[
\frac{W}{D_R}\longrightarrow e.
\]
\end{lemma}

\begin{proof}
Put $d=\lfloor\log_2R\rfloor$. Recall that $\tau_j=1$ for $j<H$ and $0<\tau_j\le1$ for
every $j<K$. Shifting the index range in $\sum_{\ell=0}^{K-1}\tau_\ell$ one step to the
left adds a term equal to one and removes a term between zero and one. Over $d$ such
shifts,
\[
0
\le
D_R-\sum_{\ell=0}^{K-1}\tau_\ell
=
\sum_{q=0}^{d-1}\bigl(1-\tau_{K-1-q}\bigr)
\le d.
\]
Since $\sum_{\ell=0}^{K-1}\tau_\ell=H+H\sum_{i=H}^{K-1}1/i$, we obtain
\begin{equation}
\label{eq:lower-capacity-bounds}
H+H\sum_{i=H}^{K-1}\frac1i
\le
D_R
\le
H+d+H\sum_{i=H}^{K-1}\frac1i.
\end{equation}

To estimate $W$, the definition of $a_i$ in \eqref{eq:lower-batches} gives
\[
\frac{e(H-2)}{i+1}-2^{-i}
<
a_i2^{-i}
\le
\frac{e(H-2)}{i+1}.
\]
Summing over all batches and using $\sum_{i=H}^{K-1}2^{-i}<2^{1-H}$, we obtain
\begin{equation}
\label{eq:lower-total-work-bounds}
e(H-2)\sum_{i=H}^{K-1}\frac1{i+1}-2^{1-H}
<
W
\le
e(H-2)\sum_{i=H}^{K-1}\frac1{i+1}.
\end{equation}

Write $S_-=\sum_{i=H}^{K-1}1/(i+1)$ and $S_+=\sum_{i=H}^{K-1}1/i$. Comparing each
summand with the integral of $1/x$ over the corresponding unit interval, and recalling
that $K=H^2$, gives
\[
S_-
\le
\int_H^K\frac{dx}{x}
=
\log H
\le
S_+,
\qquad
S_+-S_-=\frac1H-\frac1K.
\]
Consequently, $S_-/\log H\to1$ and $S_+/\log H\to1$. Since $R$, and hence $d$, is fixed,
\eqref{eq:lower-capacity-bounds} and \eqref{eq:lower-total-work-bounds} imply
\[
\frac{D_R}{H\log H}\longrightarrow1,
\qquad
\frac{W}{H\log H}\longrightarrow e.
\]
Taking their quotient proves the lemma.
\end{proof}

\begin{proof}[Proof of Theorem~\ref{thm:lb-main}]
\enlargethispage{\baselineskip}
Fix $0<\kappa<e$. Choose
\[
R>\max\left\{1,\frac{\kappa e}{e-\kappa}\right\},
\]
so that $e/(1+e/R)>\kappa$. For this fixed $R$, Lemma~\ref{lem:lower-harmonic} gives
\[
\frac{W}{D_R+W/R}=\frac{W/D_R}{1+(W/D_R)/R}\to\frac{e}{1+e/R}
\]
as $H\to\infty$. We can
therefore choose a finite integer $H\ge3$ for which
\[
\frac{W}{D_R+W/R}>\kappa.
\]
This fixes the job sequence $I$ and the original $m_0=2^{H^2}-1$ machines, independently
of the online algorithm.

To obtain an instance with any integer $m\ge m_0$, add $m-m_0$ machines of speed
\[
s_{\mathrm{add}}
=
\frac{p_{\min}}{2(1+\lceil R\rceil)},
\qquad
p_{\min}=2^{-(K-1)}.
\]
These speeds are positive rationals. Every job in $I$ takes at least
\[
\frac{p_{\min}}{s_{\mathrm{add}}}
=
2(1+\lceil R\rceil)
>
\max\{1,R\}
\]
time units on an added machine. By Lemma~\ref{lem:lower-integral-schedules}, every $I_h$
admits a schedule of makespan at most one using only the original machines. Any schedule
assigning a job to an added machine has makespan greater than one. Thus adding these
machines leaves $\OPT_h$ unchanged for every $1\le h\le n$.

Fix any randomized online algorithm $\mathcal A$ on this $m$-machine speed vector. Every
job assigned to an added machine has processing time greater than $R$, so its work is
included in $V$ in the proof of Lemma~\ref{lem:lower-fixed-length}. The bounds
\eqref{eq:lower-good-work} and \eqref{eq:lower-bad-work} therefore still yield
\eqref{eq:lower-pathwise-work}, with the sum taken over the original $K$ levels. Taking
expectations as in that proof gives
\[
\max_{1\le h\le n}
\frac{\mathbb{E}[C_{\max}(h)]}{\OPT_h}
\ge
\frac{W}{D_R+W/R}
>
\kappa.
\]

Let $h$ be the least index satisfying $\mathbb{E}[C_{\max}(h)]>\kappa\OPT_h$. This index
is determined by the fixed sequence and the probability law of $\mathcal A$. Use $I_h$
as the complete input. Couple the executions of $\mathcal A$ on $I_h$ and $I$ using the
same random choices. The speed vector and the first $h$ arrivals are identical, and the
algorithm does not know the sizes or number of future jobs. The two executions therefore
make identical assignments through job $h$. Consequently,
\[
\mathbb{E}\!\left[C_{\max}^{\mathcal A}(I_h)\right]
=
\mathbb{E}[C_{\max}(h)]
>
\kappa\OPT_h
=
\kappa\OPT(I_h).
\]
The speed vector was fixed independently of $\mathcal A$, and every job size in $I_h$ is
a positive rational. Since $\kappa<e$ was arbitrary, no randomized online algorithm has
competitive ratio below $e$.
\end{proof}

Since deterministic online algorithms are a special case of randomized online
algorithms, the same asymptotic lower bound of $e$ also holds for deterministic online
algorithms.

%% file: sections/05_conclusion.tex
\section{Conclusion}\label{sec:conclusion}

We give a deterministic online algorithm with competitive ratio $\alpha_0\approx4.857178$ for makespan scheduling on related machines and establish a randomized lower bound approaching $e$ as the number of machines grows. The remaining gap is between $e$ and $\alpha_0$ for deterministic algorithms, and between $e$ and the upper bound of approximately $4.3111$ due to \citet{BermanCharikarKarpinski2000} for randomized algorithms.

%% file: references.bib
@article{FaigleKernTuran1989,
  author = {Ulrich Faigle and Walter Kern and Gy{\"o}rgy Tur{\'a}n},
  title = {On the performance of on-line algorithms for partition problems},
  journal = {Acta Cybernetica},
  volume = {9},
  number = {2},
  pages = {107--119},
  year = {1989},
  url = {https://cyber.bibl.u-szeged.hu/index.php/actcybern/article/view/3359}
}

@misc{XiaoTan2026,
  author = {Junwen Xiao and Zhiyi Tan},
  title = {Improved Lower Bounds for Online Parallel Machine Scheduling},
  year = {2026},
  eprint = {2611.32565},
  archivePrefix = {arXiv},
  primaryClass = {cs.DS},
  howpublished = {\href{https://arxiv.org/abs/2611.32565}{arXiv:2611.32565}}
}

@misc{Janke2026,
  author = {Maximilian Janke},
  title = {Lower Bounds Above {$\sqrt{3}$} for Online Makespan Minimization on Four Identical Machines},
  year = {2026},
  eprint = {2611.10440},
  archivePrefix = {arXiv},
  primaryClass = {cs.DS},
  howpublished = {\href{https://arxiv.org/abs/2611.10440}{arXiv:2611.10440}}
}

@article{AspnesEtAl1997,
  author  = {James Aspnes and Yossi Azar and Amos Fiat and Serge Plotkin and Orli Waarts},
  title   = {On-line routing of virtual circuits with applications to load balancing and machine scheduling},
  journal = {Journal of the ACM},
  volume  = {44},
  number  = {3},
  pages   = {486--504},
  year    = {1997},
  month   = may,
  doi     = {10.1145/258128.258201},
  url     = {https://www.cs.yale.edu/homes/aspnes/papers/stoc93-abstract.html},
  note    = {\href{https://doi.org/10.1145/258128.258201}{doi:\nolinkurl{10.1145/258128.258201}}}
}

@article{BermanCharikarKarpinski2000,
  author  = {Piotr Berman and Moses Charikar and Marek Karpinski},
  title   = {On-Line Load Balancing for Related Machines},
  journal = {Journal of Algorithms},
  volume  = {35},
  number  = {1},
  pages   = {108--121},
  year    = {2000},
  month   = apr,
  doi     = {10.1006/jagm.1999.1070},
  url     = {https://eccc.weizmann.ac.il/report/2000/001/download/},
  note    = {\href{https://doi.org/10.1006/jagm.1999.1070}{doi:\nolinkurl{10.1006/jagm.1999.1070}}. Author manuscript: \href{https://eccc.weizmann.ac.il/report/2000/001/download/}{ECCC report TR00-001 (2000)}}
}

@article{EbenlendrSgall2015,
  author  = {Tom{\'{a}}{\v{s}} Ebenlendr and Ji{\v{r}}{\'{i}} Sgall},
  title   = {A Lower Bound on Deterministic Online Algorithms for Scheduling on Related Machines Without Preemption},
  journal = {Theory of Computing Systems},
  volume  = {56},
  number  = {1},
  pages   = {73--81},
  year    = {2015},
  month   = jan,
  doi     = {10.1007/s00224-013-9451-6},
  url     = {https://iuuk.mff.cuni.cz/~sgall/ps/rellbdet.pdf},
  note    = {\href{https://doi.org/10.1007/s00224-013-9451-6}{doi:\nolinkurl{10.1007/s00224-013-9451-6}}}
}

@article{EpsteinSgall2000,
  author  = {Leah Epstein and Ji{\v{r}}{\'{i}} Sgall},
  title   = {A lower bound for on-line scheduling on uniformly related machines},
  journal = {Operations Research Letters},
  volume  = {26},
  number  = {1},
  pages   = {17--22},
  year    = {2000},
  month   = feb,
  doi     = {10.1016/S0167-6377(99)00062-0},
  url     = {https://iuuk.mff.cuni.cz/~sgall/ps/rel-lb.ps},
  note    = {\href{https://doi.org/10.1016/S0167-6377(99)00062-0}{doi:\nolinkurl{10.1016/S0167-6377(99)00062-0}}}
}

@article{ChoSahni1980,
  author = {Yookun Cho and Sartaj Sahni},
  title = {Bounds for List Schedules on Uniform Processors},
  journal = {SIAM Journal on Computing},
  volume = {9},
  number = {1},
  pages = {91--103},
  year = {1980},
  url = {https://epubs.siam.org/doi/abs/10.1137/0209007},
  doi = {10.1137/0209007},
  note = {\href{https://doi.org/10.1137/0209007}{doi:\nolinkurl{10.1137/0209007}}}
}

@article{EpsteinEtAl2001Two,
  author = {Leah Epstein and John Noga and Steven S. Seiden and Ji{\v{r}}{\'i} Sgall and Gerhard J. Woeginger},
  title = {Randomized on-line scheduling on two uniform machines},
  journal = {Journal of Scheduling},
  volume = {4},
  number = {2},
  pages = {71--92},
  year = {2001},
  url = {https://onlinelibrary.wiley.com/doi/abs/10.1002/jos.60},
  doi = {10.1002/jos.60},
  note = {\href{https://doi.org/10.1002/jos.60}{doi:\nolinkurl{10.1002/jos.60}}}
}

@article{JezEtAl2013,
  author = {{\L}ukasz Je{\.{z}} and Jarett Schwartz and Ji{\v{r}}{\'i} Sgall and J{\'o}zsef B{\'e}k{\'e}si},
  title = {Lower bounds for online makespan minimization on a small number of related machines},
  journal = {Journal of Scheduling},
  volume = {16},
  pages = {539--547},
  year = {2013},
  url = {https://link.springer.com/article/10.1007/s10951-012-0288-7},
  doi = {10.1007/s10951-012-0288-7},
  note = {\href{https://doi.org/10.1007/s10951-012-0288-7}{doi:\nolinkurl{10.1007/s10951-012-0288-7}}}
}

@article{Graham1966,
  author = {Ronald L. Graham},
  title = {Bounds for Certain Multiprocessing Anomalies},
  journal = {Bell System Technical Journal},
  volume = {45},
  number = {9},
  pages = {1563--1581},
  year = {1966},
  url = {https://onlinelibrary.wiley.com/doi/10.1002/j.1538-7305.1966.tb01709.x},
  doi = {10.1002/j.1538-7305.1966.tb01709.x},
  note = {\href{https://doi.org/10.1002/j.1538-7305.1966.tb01709.x}{doi:\nolinkurl{10.1002/j.1538-7305.1966.tb01709.x}}}
}

@article{BartalEtAl1995,
  author = {Yair Bartal and Amos Fiat and Howard Karloff and Rakesh Vohra},
  title = {New algorithms for an ancient scheduling problem},
  journal = {Journal of Computer and System Sciences},
  volume = {51},
  number = {3},
  pages = {359--366},
  year = {1995},
  url = {https://cris.tau.ac.il/en/publications/new-algorithms-for-an-ancient-scheduling-problem-2/},
  doi = {10.1006/jcss.1995.1074},
  note = {\href{https://doi.org/10.1006/jcss.1995.1074}{doi:\nolinkurl{10.1006/jcss.1995.1074}}}
}

@article{Albers1999,
  author = {Susanne Albers},
  title = {Better Bounds for Online Scheduling},
  journal = {SIAM Journal on Computing},
  volume = {29},
  number = {2},
  pages = {459--473},
  year = {1999},
  url = {https://www14.in.tum.de/personen/albers/papers/stoc97a.pdf},
  doi = {10.1137/S0097539797324874},
  note = {\href{https://doi.org/10.1137/S0097539797324874}{doi:\nolinkurl{10.1137/S0097539797324874}}}
}

@article{FleischerWahl2000,
  author = {Rudolf Fleischer and Michaela Wahl},
  title = {On-line scheduling revisited},
  journal = {Journal of Scheduling},
  volume = {3},
  number = {6},
  pages = {343--353},
  year = {2000},
  url = {https://onlinelibrary.wiley.com/doi/10.1002/1099-1425%28200011/12%293%3A6%3C343%3A%3AAID-JOS54%3E3.0.CO%3B2-2},
  doi = {10.1002/1099-1425(200011/12)3:6<343::AID-JOS54>3.0.CO;2-2},
  note = {\href{https://doi.org/10.1002/1099-1425(200011/12)3:6%3C343::AID-JOS54%3E3.0.CO;2-2}{doi:\nolinkurl{10.1002/1099-1425(200011/12)3:6<343::AID-JOS54>3.0.CO;2-2}}}
}

@article{KargerPhillipsTorng1996,
  author = {David R. Karger and Steven J. Phillips and Eric Torng},
  title = {A Better Algorithm for an Ancient Scheduling Problem},
  journal = {Journal of Algorithms},
  volume = {20},
  number = {2},
  pages = {400--430},
  year = {1996},
  doi = {10.1006/jagm.1996.0019},
  note = {\href{https://doi.org/10.1006/jagm.1996.0019}{doi:\nolinkurl{10.1006/jagm.1996.0019}}}
}

@phdthesis{Rudin2001,
  author = {Rudin, III, John F.},
  title = {Improved Bounds for the On-line Scheduling Problem},
  school = {The University of Texas at Dallas},
  year = {2001}
}

@article{ChenVanVlietWoeginger1994,
  author = {Bo Chen and Andr{\'e} van Vliet and Gerhard J. Woeginger},
  title = {A Lower Bound for Randomized On-line Scheduling Algorithms},
  journal = {Information Processing Letters},
  volume = {51},
  number = {5},
  pages = {219--222},
  year = {1994},
  doi = {10.1016/0020-0190(94)00110-3},
  note = {\href{https://doi.org/10.1016/0020-0190(94)00110-3}{doi:\nolinkurl{10.1016/0020-0190(94)00110-3}}}
}

@article{Sgall1997,
  author = {Ji{\v{r}}{\'{i}} Sgall},
  title = {A Lower Bound for Randomized On-line Multiprocessor Scheduling},
  journal = {Information Processing Letters},
  volume = {63},
  number = {1},
  pages = {51--55},
  year = {1997},
  doi = {10.1016/S0020-0190(97)00093-8},
  note = {\href{https://doi.org/10.1016/S0020-0190(97)00093-8}{doi:\nolinkurl{10.1016/S0020-0190(97)00093-8}}}
}

@inproceedings{Albers2002,
  author = {Susanne Albers},
  title = {On Randomized Online Scheduling},
  booktitle = {Proceedings of the Thirty-Fourth Annual ACM Symposium on Theory of Computing},
  series = {STOC 2002},
  publisher = {Association for Computing Machinery},
  pages = {134--143},
  year = {2002},
  doi = {10.1145/509907.509930},
  note = {\href{https://doi.org/10.1145/509907.509930}{doi:\nolinkurl{10.1145/509907.509930}}}
}

@article{AlbersJanke2021,
  author = {Susanne Albers and Maximilian Janke},
  title = {Scheduling in the Random-Order Model},
  journal = {Algorithmica},
  volume = {83},
  number = {9},
  pages = {2803--2832},
  year = {2021},
  doi = {10.1007/s00453-021-00841-8},
  note = {\href{https://doi.org/10.1007/s00453-021-00841-8}{doi:\nolinkurl{10.1007/s00453-021-00841-8}}}
}
